\documentclass[11pt,a4paper]{article}
\pdfoutput=1
\usepackage{jheppub}
\usepackage{tikz}
\usetikzlibrary{intersections, calc, arrows.meta, bending}
\usepackage{subcaption}
\usepackage{simpler-wick}
\DeclareMathOperator{\tr}{tr}

\newcommand{\ri}{\mathrm{i}}
\renewcommand{\th}{\theta}

\newcommand{\cob}{\delta}

\newcommand{\hf}{\frac{1}{2}}
\newcommand{\qu}{\frac{1}{4}}

\newcommand{\del}{\partial}

\newcommand{\lap}{\Delta}
\newcommand{\bra}{\langle}
\newcommand{\ket}{\rangle}
\newcommand{\la}{\lambda}

\newcommand{\h}[1]{\widehat{#1}}
\newcommand{\bt}{\beta}

\newcommand{\Ga}{\Gamma}

\newcommand{\rt}[1]{\sqrt{#1}}
\newcommand{\cO}{\mathcal{O}}

\newcommand{\cG}{\mathcal{G}}

\newcommand{\cB}{\mathcal{B}}

\newcommand{\cK}{\mathcal{K}}

\newcommand{\qb}[1]{\Bigl[~\begin{matrix}#1\end{matrix}~\Bigr]_q}

\newcommand{\fb}{\mathfrak{b}}
\makeatletter
\gdef\@fpheader{}
\makeatother
\begin{document}
\title{End of the world brane in double scaled SYK}

\author{Kazumi Okuyama}

\affiliation{Department of Physics, 
Shinshu University, 3-1-1 Asahi, Matsumoto 390-8621, Japan}

\emailAdd{kazumi@azusa.shinshu-u.ac.jp}

\abstract{
We study the end of the world (EOW) brane
in double scaled SYK (DSSYK) model.
We find that the boundary state of EOW brane is a coherent
state of the $q$-deformed oscillators
and the associated orthogonal polynomial is the continuous
big $q$-Hermite polynomial.
In a certain scaling limit, the 
big $q$-Hermite polynomial reduces to the Whittaker function,
which reproduces the wavefunction of JT gravity with an EOW brane.
We also compute the half-wormhole amplitude in DSSYK and show that
the amplitude is decomposed into the trumpet and the factor coming from
the EOW brane. 
}

\maketitle

\section{Introduction}
The double scaled SYK (DSSYK) model \cite{Cotler:2016fpe,Berkooz:2018jqr}
is a useful toy model for the study of quantum gravity and holography.
As shown in \cite{Berkooz:2018jqr}, DSSYK is exactly solvable
by using the technique of the chord diagrams and the transfer matrix.
DSSYK allows us to explore the high energy regime of holography
and quantum gravity, beyond the realm of
JT gravity and the Schwarzian modes at low energy.
The bulk dual of DSSYK has a peculiar feature:
the length scale is quantized in units of the coupling $\la$ 
and the geodesic length is replaced by the discrete chord number $n$ \cite{Lin:2022rbf}.
In \cite{Jafferis:2022wez}, it is also suggested that
the length of geodesic loop $b$ is replaced by 
a discrete length $\fb$ as well.

To understand the bulk dual of DSSYK better, in this paper we consider the
end of the world (EOW) brane in DSSYK.
EOW branes in JT gravity were very useful for the 
study of black hole information problem and the Page curve of Hawking radiation
\cite{Penington:2019kki}.
If we quantize JT gravity in the presence of 
an EOW brane, the problem boils down to a quantum mechanical problem of a
particle moving in the Morse potential, 
and the wavefunction is given by the Whittaker function \cite{Gao:2021uro}.
We will show that the DSSYK analogue of the Whittaker function
is the continuous big $q$-Hermite polynomial
$H_n(x,a|q)$, where $a$ is related to
the tension $\mu$ of EOW brane by
\begin{equation}
\begin{aligned}
 a=q^{\mu+\hf}.
\end{aligned} 
\end{equation}
We also find that the boundary state $|B_a\ket$ of the EOW brane
is a coherent state of the $q$-deformed oscillator $A_\pm$
\begin{equation}
\begin{aligned}
 A_{-}|B_a\ket=a|B_a\ket,
\end{aligned} 
\end{equation}
where $A_\pm$ acts on the chord number state $|n\ket$ 
as a creation and annihilation operator of chords.
We find that the JT gravity amplitude of the half-wormhole ending on the EOW brane 
\cite{Gao:2021uro} has a complete parallel in DSSYK.
We find that the half-wormhole amplitude in DSSYK is decomposed into
the trumpet and the factor $\frac{a^\fb}{1-q^\fb}$ coming from the EOW brane 
\begin{equation}
\begin{tikzpicture}[scale=0.7]
\draw (-2,2) arc [start angle=90,end angle=270, x radius=1, y radius=2];
\draw[dashed] (-2,2) arc [start angle=90,end angle=-90, x radius=1, y radius=2];
\draw[thick,blue] (3.25,0) arc [start angle=360,end angle=0, x radius=0.25, y radius=0.5];
\draw (-1.86,1.99) to [out=-32,in=180] (3,0.5);
\draw (-1.97,-2.02) to [out=31,in=180] (3,-0.5);
\draw (-2,-2.1) node [below]{$\bt$};
\draw (3,-0.5) node [below]{EOW};
\draw (4.3,0) node [right]{$=$};
\draw (8,2) arc [start angle=90,end angle=270, x radius=1, y radius=2];
\draw[dashed] (8,2) arc [start angle=90,end angle=-90, x radius=1, y radius=2];
\draw[thick,red] (13.25,0) arc [start angle=360,end angle=0, x radius=0.25, y radius=0.5];
\draw (8.14,1.99) to [out=-32,in=180] (13,0.5);
\draw (8.03,-2.02) to [out=31,in=180] (13,-0.5);
\draw (8,-2.1) node [below]{$\bt$};
\draw (13,-0.5) node [below]{$\fb$};
\draw (11.7,0) node [left]{$Z_{\text{trumpet}}$};
\draw (13.2,0) node [right]{{\LARGE $\times~\frac{a^\fb}{1-q^\fb}$,}};
\draw (6.7,0) node [left]{{\Large $\displaystyle \sum_{\fb=0}^\infty$}};
\end{tikzpicture}
\label{eq:trumpet-EOW}
\end{equation}
where the trumpet partition function is given by
\begin{equation}
\begin{tikzpicture}[scale=0.75]
\draw (0,2) arc [start angle=90,end angle=270, x radius=1, y radius=2];
\draw[dashed] (0,2) arc [start angle=90,end angle=-90, x radius=1, y radius=2];
%\draw[->,>=stealth, thick, blue] (5,0) circle [x radius=0.25, y radius=0.5];
\draw[thick,red] (5.25,0) arc [start angle=360,end angle=0, x radius=0.25, y radius=0.5];
%\draw[->,>=stealth, thick, blue] (5.25,0.01)--(5.25,-0.01);
\draw (0.14,1.99) to [out=-32,in=180] (5,0.5);
\draw (0.03,-2.02) to [out=31,in=180] (5,-0.5);
\draw (0,-2.1) node [below]{$\bt$};
\draw (5,-0.5) node [below]{$\fb$};
\draw (3.5,0) node [left]{$Z_{\text{trumpet}}$};
\draw (5.8,0) node [right]{$\displaystyle =~~\int_0^\pi\frac{d\th}{2\pi}e^{-\bt E(\th)}2\cos(\fb\th)$.};
\end{tikzpicture}
\label{eq:trumpet-fig}
\end{equation}
Note that the integral over the length $b$ of geodesic loop
is replaced by the sum over the discrete length $\fb$ in 
\eqref{eq:trumpet-EOW}.

This paper is organized as follows.
In section \ref{sec:review}, we review the 
transfer matrix formalism of DSSYK.
In section \ref{sec:JT}, we review 
the EOW brane in JT gravity and the computation of 
the half-wormhole amplitude in \cite{Gao:2021uro}.
In section \ref{sec:bigq}, we consider the EOW brane in DSSYK.
We find that the wavefunction in the presence of EOW brane is given by
the big $q$-Hermite polynomial. We also find that
the boundary state of EOW brane is a coherent state
of the $q$-deformed oscillators.
In section \ref{sec:wormhole},
we compute the half-wormhole amplitude in DSSYK.
Along the way, we find the trumpet partition function in DSSYK.
In section \ref{sec:trumpet}, we study the cylinder amplitude in DSSYK, which is
obtained by gluing two trumpets.
Finally we conclude in section \ref{sec:conclusion}.
In appendix \ref{app:formula} we summarize useful formulae
of the $q$-Pochhammer symbols.
In appendix \ref{app:bigq} we summarize useful
properties of the continuous big $q$-Hermite polynomial.

\section{Review of DSSYK}\label{sec:review}
We first review the result of DSSYK \cite{Berkooz:2018jqr}.
SYK model is defined by the Hamiltonian for 
$N$ Majorana fermions $\psi_i~(i=1,\cdots,N)$
with all-to-all $p$-body interaction
\begin{equation}
\begin{aligned}
 H=\ri^{p/2}\sum_{1\leq i_1<\cdots<i_p\leq N}
J_{i_1\cdots i_p}\psi_{i_1}\cdots\psi_{i_p},
\end{aligned} 
\end{equation}
where $J_{i_1\cdots i_p}$ is a random coupling drawn from the Gaussian distribution.
DSSYK is defined by the scaling limit
\begin{equation}
\begin{aligned}
 N,p\to\infty\quad\text{with}\quad\la=\frac{2p^2}{N}:\text{fixed}.
\end{aligned} 
\label{eq:scaling}
\end{equation}
SYK model is exactly solvable in this double scaling limit.
As shown in \cite{Berkooz:2018jqr}, the ensemble average of the moment $\tr H^k$ 
reduces to a counting problem
of the intersection number of chord diagrams
\begin{equation}
\begin{aligned}
 \bra \tr H^k\ket_J=\sum_{\text{chord diagrams}}q^{\#(\text{intersections})},
\end{aligned} 
\end{equation}
where $q$ is given by
\begin{equation}
\begin{aligned}
 q=e^{-\la}.
\end{aligned} 
\end{equation}

One can introduce
a matter field in the bulk which is dual to an
operator in DSSYK. One simple example 
is the length $s$ strings of Majorana fermions
\begin{equation}
\begin{aligned}
 \cO_\lap=\ri^{s/2}\sum_{i_1\cdots i_s}K_{i_1\cdots i_s}\psi_{i_1}\cdots\psi_{i_s}
\end{aligned} 
\label{eq:matter-M}
\end{equation}
with Gaussian random coefficients $K_{i_1\cdots i_s}$
which is drawn independently from the random coupling
$J_{i_1\cdots i_p}$ in the SYK Hamiltonian.
The effect of this operator can be made finite by taking the scaling limit
\begin{equation}
\begin{aligned}
 p,s\to\infty\quad\text{with}\quad\lap=\frac{s}{p}:\text{fixed}.
\end{aligned} 
\end{equation}
In this limit, the random average of the correlator
$\tr(e^{-\bt_2 H}\cO_\lap e^{-\bt_1 H}\cO_\lap)$ can be computed 
using the technique of the chord diagram
\begin{equation}
\begin{aligned}
 \bra\tr(e^{-\bt_2 H}\wick{\c \cO_\lap e^{-\bt_1 H}\c \cO_\lap})\ket_J
=\sum_{\text{chord diagrams}}q^{\# (H\text{-}H~\text{intersections})}q^{\lap
\#(H\text{-}\cO~\text{intersections})}.
\end{aligned} 
\end{equation}

As shown in \cite{Berkooz:2018jqr}, the counting problem of
the chord diagram can be solved by introducing the transfer matrix
$T$
\begin{equation}
\begin{aligned}
 T=\frac{A_{-}+A_{+}}{\rt{1-q}},
\end{aligned} 
\end{equation}
where $A_{\pm}$ are the $q$-deformed oscillators
\begin{equation}
\begin{aligned}
 A_{-}A_{+}-qA_{+}A_{-}=1-q.
\end{aligned} 
\label{eq:qalg}
\end{equation}
$A_{\pm}$ acts on the states $\{|n\ket\}_{n=0,1,\cdots}$ as
\begin{equation}
\begin{aligned}
 A_-|n\ket=\rt{1-q^n}|n-1\ket,\quad A_+|n\ket=
\rt{1-q^{n+1}}|n+1\ket,
\end{aligned} 
\label{eq:Apm-act}
\end{equation}
where $n$ labels the number of chords and $A_\pm$ creates/annihilates the chord.
Then the disk partition function of DSSYK is written as
\begin{equation}
\begin{aligned}
 Z(\bt)=\bra \tr e^{-\bt H}\ket_J=\bra 0|e^{\bt T}|0\ket.
\end{aligned} 
\label{eq:Zbt}
\end{equation} 

One can diagonalize $T$ by using the $q$-Hermite polynomial 
\begin{equation}
\begin{aligned}
 \bra\th|n\ket=\frac{H_n(\cos\th|q)}{\rt{(q;q)_n}},
\end{aligned}
\label{eq:th|n} 
\end{equation}
which is orthogonal with respect to the measure $\mu(\th)$
defined by \footnote{See appendix 
\ref{app:formula} for the definition of the $q$-Pochhammer symbol.}
\begin{equation}
\begin{aligned}
 \mu(\th)=(q,e^{\pm2\ri\th};q)_\infty,
\end{aligned} 
\end{equation}
and the orthogonality relation reads
\begin{equation}
\begin{aligned}
 \int_0^\pi\frac{d\th}{2\pi}\mu(\th)\bra n|\th\ket\bra\th|m\ket=
\bra n|m\ket=\cob_{n,m}.
\end{aligned} 
\end{equation}
Note that $|\th\ket$ is normalized as
\begin{equation}
\begin{aligned}
 \bra\th|\th'\ket=\frac{2\pi}{\mu(\th)}\cob(\th-\th'),
\end{aligned} 
\end{equation}
and the resolution of the identity is written as
\begin{equation}
\begin{aligned}
 1=\sum_{n=0}^\infty|n\ket\bra n|
=\int_0^\pi\frac{d\th}{2\pi}\mu(\th)|\th\ket\bra\th|.
\end{aligned} 
\end{equation}
The transfer matrix $T$ is diagonal in the $|\th\ket$-basis
\begin{equation}
\begin{aligned}
 T|\th\ket=-E(\th)|\th\ket,
\end{aligned} 
\end{equation}
where the eigenvalue $E(\th)$ is given by
\begin{equation}
\begin{aligned}
 E(\th)=-u_0\cos\th,\quad u_0=\frac{2}{\rt{1-q}}.
\end{aligned} 
\label{eq:E-th}
\end{equation}
One can see that the energy spectrum is supported in the range 
$E\in [-u_0,u_0]$ and $u_0$ specifies the edge of the spectrum.
Now the disk partition function \eqref{eq:Zbt} becomes
\begin{equation}
\begin{aligned}
 Z(\bt)=\int_0^\pi\frac{d\th}{2\pi}\mu(\th) e^{-\bt E(\th)}.
\end{aligned} 
\end{equation}

One can also compute the correlator of matter operators in DSSYK using
the technique of the chord diagrams and the transfer matrix $T$.
After the Wick contraction of $K_{i_1,\cdots,i_s}$ in \eqref{eq:matter-M}, the bi-local operator
$\cB_{\bt,\lap}=\wick{\c \cO_\lap e^{-\bt H}\c \cO_\lap}$ becomes \cite{Berkooz:2018jqr}
\begin{equation}
\begin{aligned}
 \cB_{\bt,\lap}&=\sum_{n,m,i=0}^\infty (q^{2\lap};q)_i
\rt{\qb{n+i\\i}\qb{m+i\\i}}
|m+i\ket\bra m| q^{\lap \h{N}}e^{\bt T}q^{\lap \h{N}}|n\ket
\bra n+i|,
\end{aligned} 
\end{equation} 
where $\h{N}$ is the number operator
\begin{equation}
\begin{aligned}
 \h{N}|n\ket=n|n\ket.
\end{aligned} 
\end{equation}
As shown in \cite{Berkooz:2018jqr}, $\cB_{\bt,\lap}$ is invariant
under the time translation and commutes with $T$
\begin{equation}
\begin{aligned}
 \bigl[ T,\cB_{\bt,\lap}\bigr]=0.
\end{aligned} 
\end{equation}
This implies that $T$ and $\cB_{\bt,\lap}$ are simultaneously diagonalized in the $|\th\ket$-basis and the eigenvalue of $\cB_{\bt,\lap}$ is given by
\begin{equation}
\begin{aligned}
 \cB_{\bt,\lap}|\th\ket&=|\th\ket\bra\th|q^{\lap \h{N}}e^{\bt T}|0\ket
&=|\th\ket \int_0^\pi\frac{d\th'}{2\pi} \bra\th|q^{\lap \h{N}}|\th'\ket e^{-\bt E(\th')},
\end{aligned} 
\label{eq:B-eigen}
\end{equation}
where the factor $\bra\th|q^{\lap \h{N}}|\th'\ket$
in \eqref{eq:B-eigen} is given by
\begin{equation}
\begin{aligned}
 \bra\th|q^{\lap \h{N}}|\th'\ket=
\frac{(q^{2\lap};q)_\infty}{(q^{\lap}e^{\ri(\pm\th\pm\th')};q)_\infty}.
\end{aligned} 
\end{equation}
This factor also appears in the two-point function of matter operators $\cO_\lap$
\begin{equation}
\begin{aligned}
\bra 0|e^{\bt_1 T}\cB_{\bt_2,\lap}|0\ket
=\bra 0|e^{\bt_1 T}q^{\lap \h{N}}e^{\bt_2 T}|0\ket
=\int\prod_{i=1,2}\frac{d\th_i}{2\pi}e^{-\bt_i E(\th_i)}
\bra \th_1|q^{\lap \h{N}}|\th_2\ket.
\end{aligned} 
\label{eq:2pt}
\end{equation}
See also \cite{Goel:2023svz,Mukhametzhanov:2023tcg,Okuyama:2023bch}
for the semi-classical limit of matter correlators in DSSYK.

\section{EOW brane in JT gravity}\label{sec:JT}
In this section, we review the wavefunction of JT gravity in the presence of an EOW
brane \cite{Gao:2021uro}.
As shown in \cite{Gao:2021uro}, quantization of
JT gravity with an EOW brane leads to the Hamiltonian
with Morse potential and the eigenvalue equation for this Hamiltonian reads
\begin{equation}
\begin{aligned}
 -k^2\Psi=\left[\del_L^2-\mu e^{-L}-\qu e^{-2L}\right]\Psi,
\end{aligned} 
\label{eq:wave-eow}
\end{equation}
where $L$ denotes the renormalized geodesic length between 
the EOW brane and the $AdS_2$ boundary where SYK lives, 
and $\mu$ is the tension of the brane. 
Note that $L$ can be negative since $L$ is a renormalized length.
One can show that \eqref{eq:wave-eow} is solved by the Whittaker function
\begin{equation}
\begin{aligned}
 \Psi_k(L)=e^{\hf L}W_{-\mu,\ri k}(e^{-L}),
\end{aligned} 
\label{eq:Whittaker}
\end{equation}
which is normalized as
\begin{equation}
\begin{aligned}
\int_{-\infty}^\infty dL\frac{\rho(k)}{2\pi}|\Ga(\mu+1/2+\ri k)|^2 \Psi_{k}(L)\Psi_{k'}(L)&=
\cob(k-k'),\\
\int_0^\infty \frac{dk}{2\pi} 
\rho(k)|\Ga(\mu+1/2+\ri k)|^2 \Psi_{k}(L)\Psi_{k}(L')&=\cob( L- L'),
\end{aligned} 
\label{eq:ortho-W}
\end{equation}
where $\rho(k)$ is the density of states of the Schwarzian theory 
\cite{Stanford:2017thb}
\begin{equation}
\begin{aligned}
 \rho(k)=\frac{2k\sinh(2\pi k)}{\pi}.
\end{aligned} 
\label{eq:rho}
\end{equation}

As discussed in \cite{Gao:2021uro}, one can extract the coupling between
the trumpet and the EOW brane
by considering a
half-wormhole ending on the EOW brane.
The half-wormhole amplitude is defined by
\begin{equation}
\begin{aligned}
 Z_{\text{half-wormhole}}(\bt,\mu)=
\int_{-\infty}^\infty dL\int_0^\infty \frac{dk}{2\pi} 
\rho(k)|\Ga(\mu+1/2+\ri k)|^2 \Psi_{k}(L)e^{-\bt k^2}\Psi_{k}(L),
\end{aligned} 
\label{eq:JT-wormhole}
\end{equation}
which is schematically depicted as
\begin{equation}
\begin{tikzpicture}[scale=0.75]
\draw (0,-2)--(0,2);
\draw[thick, blue]  (4,-1)--(4,1);
\draw[thick,orange] (0,2)--(4,1);
\draw[thick,orange] (0,-2)--(4,-1);
\draw (5.5,0) node {$=$};
\draw (-0.2,0) node [left]{$\displaystyle \int_{-\infty}^\infty dL$};
\draw (2,1.5) node [above]{$L$};
\draw (2,-1.5) node [below]{$L$};
\draw (4.2,-1.2) node [below]{EOW};
\draw (8,2) arc [start angle=90,end angle=270, x radius=1, y radius=2];
\draw[dashed] (8,2) arc [start angle=90,end angle=-90, x radius=1, y radius=2];
%\draw[->,>=stealth, thick, blue] (5,0) circle [x radius=0.25, y radius=0.5];
\draw[thick, blue] (13.25,0) arc [start angle=360,end angle=0, x radius=0.25, y radius=0.5];
%\draw[->,>=stealth, thick, blue] (5.25,0.01)--(5.25,-0.01);
\draw (8.14,1.99) to [out=-32,in=180] (13,0.5);
\draw (8.03,-2.02) to [out=31,in=180] (13,-0.5);
\draw (8,-2.1) node [below]{$\bt$};
\draw (13.2,-0.6) node [below]{EOW};
\draw (0,0) node [right]{$\bt$};
\end{tikzpicture}
\label{eq:half-wormhole1}
\end{equation}
The integral over $L$ represents a trace on the Hilbert space spanned
by the states $\{\Psi_k(L)\}$
and the top and the bottom of the left hand side of \eqref{eq:half-wormhole1}
are periodically identified.
Thus the corresponding spacetime is a half-wormhole as shown on
the right hand side of \eqref{eq:half-wormhole1}.
This trace is obviously divergent due to the first relation in \eqref{eq:ortho-W}.
However, as demonstrated in \cite{Gao:2021uro},
one can rewrite this amplitude as an integral over the length 
$b$ of geodesic loop and interpret this divergence as a UV divergence coming from
$b=0$. 

Here we review the computation in \cite{Gao:2021uro}, which serves as a warm-up
for a similar computation of DSSYK in section \ref{sec:wormhole}.
As we will see in section \ref{sec:wormhole}, the computation below has a complete parallel
in DSSYK.
First, we rewrite
$\Psi_k(L)^2$ appearing in \eqref{eq:JT-wormhole} 
by using the formula 6.647-1 in \cite{gradshteyn2014table}
\begin{equation}
\begin{aligned}
 |\Ga(\mu+1/2+\ri k)|^2 \Psi_k(L)^2&=
2\int_0^\infty dxx^{\mu-\hf}(2z+x)^{-\mu-\hf}e^{-x-z}
K_{2\ri k}\Bigl(\rt{x(2z+x)}\Bigr),
\end{aligned} 
\label{eq:W^2-formula}
\end{equation}
where $z=e^{-L}$
and $K_\nu(x)$ denotes the modified Bessel function of the second kind.
If we further make a change of variable
\begin{equation}
\begin{aligned}
 x=\frac{2z}{e^b-1},
\end{aligned} 
\end{equation}
then \eqref{eq:W^2-formula} becomes
\begin{equation}
\begin{aligned}
 |\Ga(\mu+1/2+\ri k)|^2 \Psi_k(L)^2&=
\int_0^\infty db \,e^{-\mu b}\cG(b,e^{-L},k),
\end{aligned} 
\label{eq:W^2-formula2}
\end{equation}
with
\begin{equation}
\begin{aligned}
 \cG(b,z,k)=\frac{1}{2\sinh\frac{b}{2}}2K_{2\ri k}\left(\frac{z}{\sinh\frac{b}{2}}\right)
e^{-z\coth\frac{b}{2}}.
\end{aligned} 
\end{equation}
Then the half-wormhole amplitude \eqref{eq:JT-wormhole} is written as
\begin{equation}
\begin{aligned}
 Z_{\text{half-wormhole}}(\bt,\mu)=
\int_{-\infty}^\infty dL\int_0^\infty db\, e^{-\mu b}\cK_\bt(b,e^{-L})
\end{aligned} 
\end{equation}
where $\cK_\bt(b,z)$ is the so-called 
boundary particle propagator \cite{Suh:2020lco,Yang:2018gdb,Kitaev:2018wpr}\footnote{See also (2.42) in \cite{Kolchmeyer:2023gwa}.
Our $\cK_\bt(b,e^{-w})$ corresponds to
$\cK^t_\bt(b,w|0,w)$ in \cite{Kolchmeyer:2023gwa}.}
\begin{equation}
\begin{aligned}
 \mathcal{K}_\bt(b,z)=
\int_0^\infty \frac{dk}{2\pi}\rho(k)e^{-\bt k^2}
\cG(b,z,k).
\end{aligned} 
\label{eq:Kbt-def}
\end{equation}
The integral over $L$ can be 
performed by using the integration formula of the modified Bessel function
\begin{equation}
\begin{aligned}
 \int_0^\infty \frac{dx}{x}2K_{2\ri k}(x)e^{-x\cosh\frac{b}{2}}=
\frac{2\cos(bk)}{\rho(k)},
\end{aligned}
\label{eq:martinec} 
\end{equation}
which was heavily
utilized in the study of minimal string theory \cite{Martinec:2003ka}. 
Using this formula, we find
\begin{equation}
\begin{aligned}
 \int_{-\infty}^\infty dL\, \cG(b,e^{-L},k)=
\frac{1}{2\sinh\frac{b}{2}}\frac{2\cos(bk)}{\rho(k)}.
\end{aligned} 
\label{eq:K-int}
\end{equation}
Using \eqref{eq:K-int}, the $L$ integral of the boundary particle propagator
in \eqref{eq:Kbt-def} becomes \footnote{The fact that the $L$ integral of
the boundary particle propagator $\mathcal{K}_\bt(b,e^{-L})$
is proportional to the trumpet was also pointed out in \cite{Kolchmeyer:2023gwa}.}
\begin{equation}
\begin{aligned}
 \int_{-\infty}^\infty dL\mathcal{K}_\bt(b,e^{-L})
=\frac{1}{2\sinh\frac{b}{2}}Z_{\text{trumpet}}(\bt,b),
\end{aligned} 
\label{eq:Kbt-int}
\end{equation}
where the trumpet partition function is given by
\begin{equation}
\begin{aligned}
 Z_{\text{trumpet}}(\bt,b)&=\int_0^\infty \frac{dk}{2\pi}e^{-\bt k^2}2\cos(bk)
=\frac{e^{-\frac{b^2}{4\bt}}}{\rt{4\pi\bt}}.
\end{aligned} 
\end{equation}
This agrees with the known result of trumpet in JT gravity \cite{Saad:2019lba},
as expected. Note that
$\rho(k)$ is canceled in this computation and the trumpet is independent of
$\rho(k)$.
Finally, we arrive at the desired expression of the half-wormhole amplitude
\begin{equation}
\begin{aligned}
Z_{\text{half-wormhole}}(\bt,\mu)=
\int_0^\infty db \frac{e^{-\mu b}}{2\sinh\frac{b}{2}}
Z_{\text{trumpet}}(\bt,b).
\end{aligned}
\label{eq:EOWampJT}
\end{equation} 
This amplitude 
has a divergence coming from $b=0$, which is interpreted as a
UV divergence in the bulk 
gravitational theory.

\section{EOW brane in DSSYK}\label{sec:bigq}
In this section, we consider the EOW brane in DSSYK.
It turns out that 
the big $q$-Hermite polynomial 
$H_n(x,a|q)$ plays the role of the 
Whittaker function in \eqref{eq:Whittaker}.

\subsection{Triple scaling limit}
The continuous big $q$-Hermite polynomial $H_n(x,a|q)$ is a 
one parameter generalization of the $q$-Hermite polynomial $H_n(x|q)$.
$H_n(x,a|q)$ 
reduces to $H_n(x|q)$ when $a=0$. 
$H_n(x,a|q)$ is defined recursively 
by the three-term recursion relation
\begin{equation}
\begin{aligned}
 2x H_n(x,a|q)=H_{n+1}(x,a|q)+aq^nH_n(x,a|q)+(1-q^n)H_{n-1}(x,a|q), 
\end{aligned} 
\label{eq:bigq-rec}
\end{equation}
with the initial condition $H_{-1}(x,a|q)=0,H_0(x,a|q)=1$.
An important property of $H_n(x,a|q)$
is the orthogonality relation
\begin{equation}
\begin{aligned}
 \int_0^\pi\frac{d\th}{2\pi}\mu(\th)\frac{1}{(ae^{\pm\ri\th};q)_\infty}
H_n(\cos\th,a|q)H_m(\cos\th,a|q)=(q;q)_n\cob_{n,m}.
\end{aligned} 
\label{eq:ortho-big}
\end{equation}
It is convenient to introduce the  normalized wavefunction
\begin{equation}
\begin{aligned}
 \psi_n(\th)=\frac{H_n(\cos\th,a|q)}{\rt{(q;q)_n}},
\end{aligned} 
\label{eq:psi-n}
\end{equation}
which satisfies the orthogonality relation
\begin{equation}
\begin{aligned}
 \int_0^\pi\frac{d\th}{2\pi}\mu(\th)\frac{1}{(ae^{\pm\ri\th};q)_\infty}
\psi_n(\th)\psi_m(\th)=\cob_{n,m}.
\end{aligned} 
\end{equation}

We will show that $\psi_n(\th)$
is the DSSYK analogue of the wavefunction $\Psi_k(L)$ \eqref{eq:Whittaker}
of JT gravity with an EOW brane.
In terms of the wavefunction $\psi_n(\th)$
in \eqref{eq:psi-n}, the three-term recursion relation \eqref{eq:bigq-rec} 
is written as
\begin{equation}
\begin{aligned}
 2\cos\th\psi_n(\th)=\rt{1-q^{n+1}}\psi_{n+1}(\th)+aq^n\psi_n(\th)+
\rt{1-q^n}\psi_{n-1}(\th).
\end{aligned}
\label{eq:psin-rec} 
\end{equation}
To see the semi-classical picture of \eqref{eq:psin-rec}, 
we consider the triple scaling limit
\begin{equation}
\begin{aligned}
 \la\to0,\quad a=q^{\mu+\hf}, \quad q^n=\la e^{-L},\quad \th=\la k,
\end{aligned} 
\label{eq:triple}
\end{equation}
with $\mu,L,$ and $k$ fixed.
If we denote $\psi_n(\th)=\Psi_k(L)$ \footnote{Readers should not be confused by the fact that the argument and the subscript
are exchanged between $\psi_n(\th)$ and $\Psi_k(L)$.}
in this limit, the right hand side of
\eqref{eq:psin-rec} becomes
\begin{equation}
\begin{aligned}
 &\rt{1-\la e^{-L-\la}}\Psi_k(L+\la)+\la e^{-\la(\mu+1/2)-L}\Psi_k(L)
+\rt{1-\la e^{-L}}\Psi_k(L-\la)\\
=&\,2\Psi_k(L)+\la^2\left[\del_L^2-\mu e^{-L}-\qu e^{-2L}\right]\Psi_k(L)+\cO(\la^3).
\end{aligned} 
\end{equation}
Similarly, the left hand side of \eqref{eq:psin-rec} is expanded as
\begin{equation}
\begin{aligned}
 2\cos(\la k)\Psi_k(L)=2\Psi_k(L)-\la^2k^2\Psi_k(L)+\cO(\la^3).
\end{aligned} 
\end{equation}
Finally, equating the $\cO(\la^2)$ term of the both sides of \eqref{eq:psin-rec}
we find 
\begin{equation}
\begin{aligned}
 -k^2\Psi_k(L)=\left[\del_L^2-\mu e^{-L}-\qu e^{-2L}\right]\Psi_k(L),
\end{aligned}
\label{eq:wave-eq} 
\end{equation}
which is exactly the 
equation \eqref{eq:wave-eow} for JT gravity with an EOW brane!
This strongly suggests that $\psi_n(\th)$ in \eqref{eq:psi-n} is
the DSSYK analogue of the EOW wavefunction $\Psi_k(L)$.
For the small $a$ expansion to be well-defined,
from \eqref{eq:triple} we see that $\mu$ should satisfy
the inequality $\mu>-\hf$. This inequality
also appears in JT gravity with an EOW brane \cite{Gao:2021uro}. 

As another consistency check, let us 
see that the measure factor correctly reduces to that of
the Whittaker function.
As shown in \cite{Berkooz:2018jqr}, in the scaling limit
\eqref{eq:triple}, $\mu(\th)$ reduces 
to the Schwarzian density of states $\rho(k)$ in \eqref{eq:rho} 
\begin{equation}
\begin{aligned}
 \mu(\th)~\to~\rho(k).
\end{aligned} 
\end{equation}
There is an 
extra factor $1/(ae^{\pm\ri\th};q)_\infty$ 
in \eqref{eq:ortho-big}, which becomes 
\begin{equation}
\begin{aligned}
 \frac{1}{(ae^{\pm\ri\th};q)_\infty}=\frac{1}{(q^{\mu+1/2\pm\ri k};q)_\infty}.
\end{aligned} 
\label{eq:bigq-mes}
\end{equation}
We observe that \eqref{eq:bigq-mes} is proportional to the
$q$-Gamma function
\begin{equation}
\begin{aligned}
 \Ga_q(z)=(1-q)^{1-z}\frac{(q;q)_\infty}{(q^z;q)_\infty}.
\end{aligned} 
\end{equation}
Since $\Ga_q(z)$ reduces to the ordinary Gamma function 
$\Ga(z)$ in the limit $q\to1$,
the scaling limit of \eqref{eq:bigq-mes} becomes
\begin{equation}
\begin{aligned}
 \frac{1}{(ae^{\pm\ri\th};q)_\infty}~\to~|\Ga(\mu+1/2+\ri k)|^2,
\end{aligned} 
\end{equation}
up to an overall factor.
This correctly
reproduces the measure factor of Whittaker function in \eqref{eq:ortho-W}.

\subsection{Boundary state of EOW brane}
Let us consider a physical interpretation of the extra measure
factor \eqref{eq:bigq-mes}.
We propose that the extra factor \eqref{eq:bigq-mes}
is the $\th$-representation of the boundary state
$|B_a\ket$ of the EOW brane
\begin{equation}
\begin{aligned}
 \frac{1}{(ae^{\pm\ri\th};q)_\infty}=\bra\th|B_a\ket.
\end{aligned} 
\label{eq:extra-factor}
\end{equation}
Using the formula \eqref{eq:H-gen}, the left hand side of 
\eqref{eq:extra-factor} is expanded as
\begin{equation}
\begin{aligned}
\frac{1}{(ae^{\pm\ri\th};q)_\infty}= \sum_{n=0}^\infty\frac{a^n}{(q;q)_n}H_n(\cos\th|q)=\sum_{n=0}^\infty\frac{a^n}{\rt{(q;q)_n}}\bra\th|n\ket,
\end{aligned} 
\label{eq:th-Ba}
\end{equation}
where we used \eqref{eq:th|n} in the last equality.
From \eqref{eq:extra-factor} and \eqref{eq:th-Ba}, 
$|B_a\ket$ is given by
\begin{equation}
\begin{aligned}
 |B_a\ket&=\sum_{n=0}^\infty\frac{a^n}{\rt{(q;q)_n}}|n\ket.
\end{aligned}
\label{eq:Ba} 
\end{equation}
Then the partition function in the presence of the extra
factor $\bra\th|B_a\ket$ is written as
\begin{equation}
\begin{aligned}
 \int_0^\pi\frac{d\th}{2\pi}\mu(\th)\bra\th|B_a\ket e^{-\bt E(\th)}
=\bra 0|e^{\bt T}|B_a\ket.
\end{aligned} 
\label{eq:half-disk2}
\end{equation}
This expression suggests that $|B_a\ket$ is
the boundary state 
of EOW brane 
and \eqref{eq:half-disk2} is the partition function
of half-disk
ending on the EOW brane
\begin{equation}
\begin{aligned}
 \begin{tikzpicture}[scale=0.75]
\draw (0,2) arc [start angle=90,end angle=270,radius=2]; 
\draw[blue,thick] (0,-2)--(0,2); 
\draw (-2.3,0) node [left]{$\bra 0|e^{\bt T}|B_a\ket=$};
\draw (0,0.1) node [right]{EOW};
\draw (-1.5,1.4) node [left]{$\bt$};
\end{tikzpicture}
\end{aligned}
\label{eq:half-disk}
\end{equation}
where the blue thick line represents the worldline of the EOW brane.

The boundary state $|B_a\ket$ in \eqref{eq:Ba} has an interesting characterization:
it 
is a coherent state of the $q$-deformed oscillator
\begin{equation}
\begin{aligned}
 A_{-}|B_a\ket=a|B_a\ket,
\end{aligned} 
\end{equation}
which can be shown by using the first relation of \eqref{eq:Apm-act}.
$|B_a\ket$ 
 is also written as
\begin{equation}
\begin{aligned}
 |B_a\ket&=\sum_{n=0}^\infty\frac{a^nA_+^n}{(q;q)_n}|0\ket
=\frac{1}{(aA_{+};q)_\infty}|0\ket,
\end{aligned} 
\end{equation}
where we used the relation
\begin{equation}
\begin{aligned}
 A_+^n|0\ket=\rt{(q;q)_n}|n\ket,
\end{aligned} 
\label{eq:A+^n}
\end{equation} 
and the formula \eqref{eq:q1}. 

In the $a\to0$ limit, $|B_a\ket$
reduces to the Hartle-Hawking vacuum $|0\ket$ \cite{Okuyama:2022szh}
\begin{equation}
\begin{aligned}
 \lim_{a\to0}|B_a\ket=|0\ket,
\end{aligned} 
\end{equation}
and the half-disk amplitude \eqref{eq:half-disk2} reduces to the disk amplitude
\eqref{eq:Zbt}
\begin{equation}
\begin{aligned}
 \lim_{a\to0}\bra 0|e^{\bt T}|B_a\ket=\bra 0|e^{\bt T}|0\ket.
\end{aligned} 
\end{equation}
Thus, $a=0$ corresponds to the absence of EOW brane.
In fact, the big $q$-Hermite polynomial $H_n(x,a|q)$ reduces to 
the $q$-Hermite polynomial $H_n(x|q)$ when $a=0$.
From the relation $a=q^{\mu+\hf}$, the absence of EOW brane
corresponds to $\mu=\infty$, not $\mu=0$.
This is similar to the situation considered in \cite{Berkooz:2020uly}:
a heavy object in the bulk corresponds to a projection
to the vacuum $|0\ket$.
For instance, the large dimension  limit of the
two-point function \eqref{eq:2pt} of matter operator $\cO_\lap$ becomes
\begin{equation}
\begin{aligned}
 \lim_{\lap\to\infty}\bra0|e^{\bt_1T}q^{\lap\h{N}}e^{\bt_2T}|0\ket
=\bra0|e^{\bt_1T}|0\ket\bra0|e^{\bt_2T}|0\ket.
\end{aligned} 
\label{eq:lim-2pt}
\end{equation}
Note that the right hand side is the product of two disk amplitudes.
\eqref{eq:lim-2pt} is schematically depicted as
\begin{equation}
\begin{aligned}
 \begin{tikzpicture}[scale=0.75]
\draw (0,0) circle [radius=2];
\draw[orange,thick] (0,-2)--(0,2); 
\draw[orange,fill=orange] (0,2) circle [radius=0.1];
\draw[orange,fill=orange] (0,-2) circle [radius=0.1];
\draw (0,0) node [right]{$\lap$};
\draw (-2.2,0) node [left]{$\displaystyle\lim_{\lap\to\infty}$};
\draw (2.2,0) node [right]{$=$};
\draw (4.5,0) circle [radius=1];
\draw (6.6,0) circle [radius=1];
\end{tikzpicture}
\end{aligned}\quad .
\label{eq:2pt-pinch}
\end{equation}
One might naively think that the heavy operator
corresponds to a black hole in the bulk, but this is not the case
in DSSYK: if we try to create a black hole by putting two heavy operators
on the boundary, then the bulk spacetime is pinched and 
splits into two pieces and the black hole never forms!

\subsection{Representation of the $q$-oscillator algebra}
Let us consider the action of $A_{\pm}$ on the continuous big $q$-Hermite 
polynomials. To this end, it is convenient to introduce the notation
\begin{equation}
\begin{aligned}
 H_n(\cos\th|q)=\bra \th|H_{n}\ket,\quad
H_n(\cos\th,a|q)=\bra \th|H_{n},a\ket.
\end{aligned} 
\end{equation}
The action of $A_\pm$ on the basis $|H_n\ket$ is given by
\begin{equation}
\begin{aligned}
 A_{+}|H_n\ket=|H_{n+1}\ket,\quad
A_{-}|H_n\ket=(1-q^n)|H_{n-1}\ket.
\end{aligned} 
\label{eq:Apm-act-Hn}
\end{equation}
From \eqref{eq:big-exp}, the big $q$-Hermite state 
$|H_{n},a\ket$ is written as a linear combination
of the $q$-Hermite state $|H_{k}\ket$
\begin{equation}
\begin{aligned}
 |H_n,a\ket=\sum_{k=0}^n \qb{n\\k}(-a)^k q^{\hf k(k-1)}|H_{n-k}\ket.
\end{aligned} 
\end{equation}
From \eqref{eq:Apm-act-Hn}, one can deduce the action of $A_\pm$ on the 
big $q$-Hermite basis
\begin{equation}
\begin{aligned}
 A_{+}|H_n,a\ket=|H_{n+1},a\ket+aq^n|H_n,a\ket,\quad
A_{-}|H_n,a\ket=(1-q^n)|H_{n-1},a\ket,
\end{aligned} 
\label{Apm-big}
\end{equation}
from which one 
can easily check that the $q$-oscillator algebra \eqref{eq:qalg} is realized on the big $q$-Hermite basis
\begin{equation}
\begin{aligned}
 (A_{-}A_+-qA_{+}A_{-})|H_n,a\ket=(1-q)|H_n,a\ket.
\end{aligned} 
\end{equation}
We should stress that the parameter $a$ does not deform the algebra itself.
Instead, $a$ labels a representation of the 
$q$-oscillator algebra \eqref{eq:qalg}.

From \eqref{Apm-big}, one can see that $A_{+}+A_{-}$
acts on the big $q$-Hermite basis as
\begin{equation}
\begin{aligned}
 (A_{+}+A_{-})|H_n,a\ket=
|H_{n+1},a\ket+aq^n|H_n,a\ket+(1-q^n)|H_{n-1},a\ket,
\end{aligned} 
\end{equation}
which is exactly the same combination appearing in the three-term recursion
of the big $q$-Hermite polynomial \eqref{eq:bigq-rec}.
Thus the transfer matrix $T=\frac{A_{+}+A_{-}}{\rt{1-q}}$
is diagonal on the basis of $H_n(\cos\th,a|q)$ as well,
with the same eigenvalue
$\frac{2\cos\th}{\rt{1-q}}$ as the original one without EOW brane. 
Namely, we can use the common transfer matrix $T=\frac{A_{+}+A_{-}}{\rt{1-q}}$
for both with and without the EOW brane.
The only difference is the presence of the
extra factor $\bra\th|B_a\ket$ for the case with the EOW brane
\begin{equation}
\begin{aligned}
 \text{pure~DSSYK}:&\quad
\int_0^\pi\frac{d\th}{2\pi}\mu(\th)e^{-\bt E(\th)}(\cdots),\\
\text{DSSYK with EOW brane} :&\quad
\int_0^\pi\frac{d\th}{2\pi}\mu(\th)\bra\th|B_a\ket
e^{-\bt E(\th)}(\cdots).
\end{aligned} 
\end{equation}

\section{Half-wormhole in DSSYK}\label{sec:wormhole}
Let us consider the DSSYK analogue of the computation 
in section \ref{sec:JT}.
Using the dictionary
\begin{equation}
\begin{aligned}
 \int_{-\infty}^\infty dL~~&\to~~\sum_{n=0}^\infty,\\
\int_0^\infty \frac{dk}{2\pi}~~&\to~~\int_0^\pi \frac{d\th}{2\pi},\\
\rho(k)~~&\to~~\mu(\th),\\
|\Ga(\mu+1/2+\ri k)|^2~~&\to~~\bra\th|B_a\ket,\\
\Psi_k(L)~~&\to~~\psi_n(\th),\\
e^{-\bt k^2}~~&\to~~e^{-\bt E(\th)},
\end{aligned} 
\end{equation}
the half-wormhole amplitude \eqref{eq:JT-wormhole} is translated into DSSYK 
as
\begin{equation}
\begin{aligned}
 Z_{\text{half-wormhole}}(\bt,a)=\sum_{n=0}^\infty
\int_0^\pi \frac{d\th}{2\pi}\mu(\th)\bra\th|B_a\ket
\psi_n(\th)e^{-\bt E(\th)}\psi_n(\th).
\end{aligned} 
\label{eq:DS-wormhole}
\end{equation}
This is schematically depicted as
\begin{equation}
\begin{tikzpicture}[scale=0.75]
\draw (0,-2)--(0,2);
\draw[thick, blue]  (4,-1)--(4,1);
\draw[thick,orange] (0,2)--(4,1);
\draw[thick,orange] (0,-2)--(4,-1);
\draw (5.5,0) node {$=$};
\draw (-0.2,0) node [left]{$\displaystyle \sum_{n=0}^\infty $};
\draw (2,1.5) node [above]{$n$};
\draw (2,-1.5) node [below]{$n$};
\draw (0,0) node [right]{$\bt$};
\draw (4.2,-1.2) node [below]{EOW};
\draw (8,2) arc [start angle=90,end angle=270, x radius=1, y radius=2];
\draw[dashed] (8,2) arc [start angle=90,end angle=-90, x radius=1, y radius=2];
%\draw[->,>=stealth, thick, blue] (5,0) circle [x radius=0.25, y radius=0.5];
\draw[thick, blue] (13.25,0) arc [start angle=360,end angle=0, x radius=0.25, y radius=0.5];
%\draw[->,>=stealth, thick, blue] (5.25,0.01)--(5.25,-0.01);
\draw (8.14,1.99) to [out=-32,in=180] (13,0.5);
\draw (8.03,-2.02) to [out=31,in=180] (13,-0.5);
\draw (8,-2.1) node [below]{$\bt$};
\draw (13.2,-0.6) node [below]{EOW};
\end{tikzpicture}
\label{eq:half-wormhole2}
\end{equation}
The sum over $n$ in \eqref{eq:DS-wormhole} is divergent due to the relation
$\sum_n \psi_n(\theta)^2\propto \delta(\theta-\theta)$.
However, we can rewrite the amplitude \eqref{eq:DS-wormhole}
as a sum over the discrete length $\fb$ of geodesic loop,
in a similar manner as the computation 
in the JT gravity case in section \ref{sec:JT}.
In analogy with \eqref{eq:W^2-formula2}, we define 
$\cG(\fb,n,\th)$ by
\begin{equation}
\begin{aligned}
 \bra\th|B_a\ket\psi_n(\th)^2&=\sum_{\fb=0}^\infty a^\fb\mathcal{G}(\fb,n,\th).
\end{aligned} 
\end{equation}
Using the small $a$ expansion of $\bra\th|B_a\ket$ in
\eqref{eq:th-Ba} and the expansion of  big $q$-Hermite polynomial
in \eqref{eq:big-exp}, together with the linearization formula of the $q$-Hermite polynomials \eqref{eq:linearization}, after some algebra we find
\begin{equation}
\begin{aligned}
 \cG(\fb,n,\th)=\sum_{k=0}^n \frac{(q;q)_n}{(q;q)_{n-k}^2(q;q)_k}
\frac{q^{k\fb}}{(q;q)_\fb}H_{2n-2k+\fb}(\cos\th|q).
\end{aligned} 
\end{equation}
Using the DSSYK analogue of the relation \eqref{eq:martinec} \footnote{
\eqref{eq:cos-formula} can be proved  as follows.
Expanding the left hand side of 
\eqref{eq:cos-formula} as
\begin{equation}
\begin{aligned}
 \frac{2\cos(\fb\th)}{\mu(\th)}=\sum_{n=0}^\infty c_n\frac{H_n(\cos\th|q)}{(q;q)_n},
\end{aligned} 
\end{equation}
$c_n$ is obtained by using the orthogonality relation of $H_n(\cos\th|q)$
\begin{equation}
\begin{aligned}
 c_n=\int_0^\pi\frac{d\th}{2\pi}\mu(\th)\left(
\sum_{j=0}^\infty c_j\frac{H_j(\cos\th|q)}{(q;q)_j}\right)H_n(\cos\th|q)=
\int_0^\pi\frac{d\th}{2\pi}2\cos(\fb\th)H_n(\cos\th|q).
\end{aligned} 
\end{equation}
Finally, using the expansion of $H_n(\cos\th|q)$ in \eqref{eq:Hn-exp} we find 
\eqref{eq:cos-formula}. 
}
\begin{equation}
\begin{aligned}
\frac{2\cos(\fb\th)}{\mu(\th)}&=\sum_{k=0}^\infty
\frac{H_{2k+\fb}(\cos\th|q)}{(q;q)_{k}(q;q)_{k+\fb}},
\end{aligned} 
\label{eq:cos-formula}
\end{equation}
we find 
\begin{equation}
\begin{aligned}
 \sum_{n=0}^\infty \cG(\fb,n,\th)=\frac{1}{1-q^\fb}\frac{2\cos(\fb\th)}{\mu(\th)}.
\end{aligned} 
\end{equation}
This is the analogue of \eqref{eq:K-int}.
It is also natural to define the DSSYK analogue of the boundary particle 
propagator
\begin{equation}
\begin{aligned}
 \cK_\bt(\fb,n)=\int_0^\pi\frac{d\th}{2\pi}\mu(\th)e^{-\bt E(\th)}\cG(\fb,n,\th).
\end{aligned} 
\end{equation}
In analogy with \eqref{eq:Kbt-int},
the sum over $n$ of $\cK_\bt(\fb,n)$ is written as
\begin{equation}
\begin{aligned}
 \sum_{n=0}^\infty\cK_\bt(\fb,n)&
=\frac{1}{1-q^\fb}Z_{\text{trumpet}}(\bt,\fb)
\end{aligned} 
\end{equation}
where we introduced the DSSYK analogue of the trumpet
\begin{equation}
\begin{aligned}
 Z_{\text{trumpet}}(\bt,\fb)=\int_0^\pi\frac{d\th}{2\pi}
e^{-\bt E(\th)}2\cos(\fb\th).
\end{aligned} 
\label{eq:DS-trumpet}
\end{equation}
Note that in this computation $\mu(\th)$ is canceled and $Z_{\text{trumpet}}(\bt,\fb)$ is independent of $\mu(\th)$.
Finally, the half-wormhole amplitude \eqref{eq:DS-wormhole} is written as
\begin{equation}
\begin{aligned}
Z_{\text{half-wormhole}}(\bt,a)
=&\sum_{\fb=0}^\infty\frac{a^\fb}{1-q^\fb}Z_{\text{trumpet}}(\bt,\fb).
\end{aligned} 
\label{eq:EOWtrumpet-sum}
\end{equation}
See \eqref{eq:trumpet-EOW}
for a pictorial representation of \eqref{eq:EOWtrumpet-sum}.
If we set $a=q^{\mu+\hf}$, \eqref{eq:EOWtrumpet-sum} becomes
\begin{equation}
\begin{aligned}
 Z_{\text{half-wormhole}}(\bt,q^{\mu+\hf})
=&\sum_{\fb=0}^\infty\frac{q^{\mu\fb}}{q^{-\hf\fb}-q^{\hf\fb}}Z_{\text{trumpet}}(\bt,\fb),
\end{aligned} 
\label{eq:DS-result}
\end{equation}
which is the DSSYK analogue of \eqref{eq:EOWampJT}
where the integral over $b$ is replaced by the sum over $\fb$.
Note that the $\fb=0$ term in \eqref{eq:DS-result} is divergent: this is 
an analogy of the fact that
the integral in \eqref{eq:EOWampJT} has a UV divergence coming from 
$b=0$.

\section{Trumpet and cylinder in DSSYK}\label{sec:trumpet}
Plugging $E(\th)=-u_0\cos\th$ \eqref{eq:E-th} into \eqref{eq:DS-trumpet},
the trumpet of DSSYK becomes \footnote{From the usual large $N$ counting,
the genus-$g$ with $n$-boundary amplitude comes with the factor
of $e^{(2-2g-n)S_0}$, where $e^{S_0}=2^{N/2}$ is the dimension of the Hilbert space
of $N$ Majorana fermions. 
For instance, the trumpet with $(g,n)=(0,2)$ scales as $e^{0S_0}$ which is suppressed 
relative to the disk with $(g,n)=(0,1)$ of the order $e^{S_0}$.
In this paper, we have suppressed this factor $e^{(2-2g-n)S_0}$ for 
simplicity.
}
\begin{equation}
\begin{aligned}
 Z_{\text{trumpet}}(\bt,\fb)=\int_0^\pi\frac{d\th}{2\pi}e^{\bt u_0\cos\th}2\cos(\fb\th)
=I_{\fb}(\bt u_0),
\end{aligned}
\label{eq:trumpet-I} 
\end{equation}  
where $I_\nu(x)$ denotes the modified Bessel function of the first kind.
As discussed in \cite{Jafferis:2022wez}, the cylinder amplitude is obtained 
as a ``double trumpet'' by
gluing two trumpets \footnote{See also \cite{Berkooz:2020fvm} for 
the study of non-planar corrections in DSSYK.}
\begin{equation}
\begin{aligned}
  Z_{\text{cylinder}}(\bt_1,\bt_2)&=
\sum_{\fb=0}^\infty \fb Z_{\text{trumpet}}(\bt_1,\fb) Z_{\text{trumpet}}(\bt_2,\fb).
\end{aligned} 
\label{eq:cyl-sum}
\end{equation}
This is schematically depicted as
\begin{equation}
\begin{aligned}
 \begin{tikzpicture}[scale=0.75]
\draw (0,2) arc [start angle=90,end angle=270, x radius=1, y radius=2];
\draw[dashed] (0,2) arc [start angle=90,end angle=-90, x radius=1, y radius=2];
\draw[thick,red] (5.25,0) arc [start angle=360,end angle=0, x radius=0.25, y radius=0.5];
\draw (0.14,1.99) to [out=-32,in=180] (5,0.5);
\draw (0.03,-2.02) to [out=31,in=180] (5,-0.5);
\draw (0,-2.1) node [below]{$\bt_1$};
\draw (10.8,0) circle [x radius=1, y radius=2];
\draw (5.7,0.5) to [out=0,in=212] (10.6,1.97);
\draw (5.7,-0.5) to [out=0,in=149] (10.6,-1.97);
\draw (10.8,-2.1) node [below]{$\bt_2$};
%\draw[->,>=stealth, thick, blue] (5.95,0) arc [start angle=360,end angle=0, x radius=0.25, y radius=0.5];
\draw[thick,red] (5.7,0.5) arc [start angle=90,end angle=270, x radius=0.25, y radius=0.5];
\draw[thick,dashed,red] (5.7,0.5) arc [start angle=90,end angle=-90, x radius=0.25, y radius=0.5];
\draw (5.45,-0.01)--(5.45,0.01);
\draw (4.9,-0.6) node [below]{$\fb$};
\draw (5.8,-0.6) node [below]{$\fb$};
\draw (-1,0) node [left]{{\Large $\displaystyle \sum_{\fb=0}^\infty \fb\times$}};
\end{tikzpicture}
\end{aligned} 
\end{equation}
Using the relation
\begin{equation}
\begin{aligned}
 \frac{2\nu}{x}I_\nu(x)=I_{\nu-1}(x)-I_{\nu+1}(x),
\end{aligned} 
\end{equation}
we can perform the summation over $\fb$
in \eqref{eq:cyl-sum} in a closed form
\begin{equation} 
\begin{aligned}
 Z_{\text{cylinder}}(\bt_1,\bt_2)
&=\frac{\bt_1\bt_2u_0}{2(\bt_1+\bt_2)}
\Bigl[I_0(\bt_1u_0)I_1(\bt_2u_0)+
I_1(\bt_1u_0)I_0(\bt_2u_0)\Bigr].
\end{aligned}
\label{eq:Z-cyl} 
\end{equation}
Note that \eqref{eq:Z-cyl} is equal to the cylinder amplitude in the Gaussian matrix model \cite{Akemann:2001st,Giombi:2009ms,Okuyama:2018aij}. 
This is expected since the cylinder amplitude is independent of the details
of the matrix model potential and it depends only on
the endpoints $\pm u_0$ of the spectrum \cite{Ambjorn:1990ji,Brezin:1993qg}.
\footnote{As argued in \cite{Okuyama:2021eju}, trumpet in topological gravity 
is also independent of the background $\{t_k\}$.}

By using the asymptotic form of the modified
Bessel function
\begin{equation}
\begin{aligned}
 I_\nu(x)\sim \frac{e^{x}}{\rt{2\pi x}},\qquad (x\gg1)
\end{aligned} 
\end{equation}
the low temperature limit of $Z_{\text{cylinder}}(\bt_1,\bt_2)$ 
becomes
\begin{equation}
\begin{aligned}
 Z_{\text{cylinder}}(\bt_1,\bt_2)\sim \frac{\rt{\bt_1\bt_2}}{2\pi(\bt_1+\bt_2)}
e^{(\bt_1+\bt_2)u_0}.
\end{aligned} 
\end{equation}
This agrees with the known result of cylinder amplitude in JT gravity 
\cite{Saad:2019lba}, up to the factor
$e^{(\bt_1+\bt_2)u_0}$ coming from the threshold energy $E=-u_0$.

In the semi-classical limit $\la\to0$ with 
\begin{equation}
\begin{aligned}
 \th=\la k,\quad b=\la \fb,\quad\bt\to\bt\la^{-\frac{3}{2}}
\end{aligned} 
\label{eq:semi}
\end{equation} 
the trumpet in \eqref{eq:trumpet-I} reduces to 
\begin{equation}
\begin{aligned}
 Z_{\text{trumpet}}(\bt,b)= \int_0^\infty \frac{dk}{2\pi} e^{-\bt k^2}2\cos(bk),
\end{aligned} 
\end{equation}
up to an overall factor. This reproduces the known result of trumpet
in JT gravity, as expected.
In the semi-classical limit \eqref{eq:semi}, the sum over $\fb$ in 
\eqref{eq:cyl-sum} is replaced by the integral over $b$
\begin{equation}
\begin{aligned}
  Z_{\text{cylinder}}(\bt_1,\bt_2)&=
\int_0^\infty db b  Z_{\text{trumpet}}(\bt_1,b) Z_{\text{trumpet}}(\bt_2,b),
\end{aligned} 
\end{equation}
which is also consistent with the JT gravity result.

\begin{figure}[t]
\centering
\subcaptionbox{$\lap=2$\label{sfig:S1}}{\includegraphics
[width=0.45\linewidth]{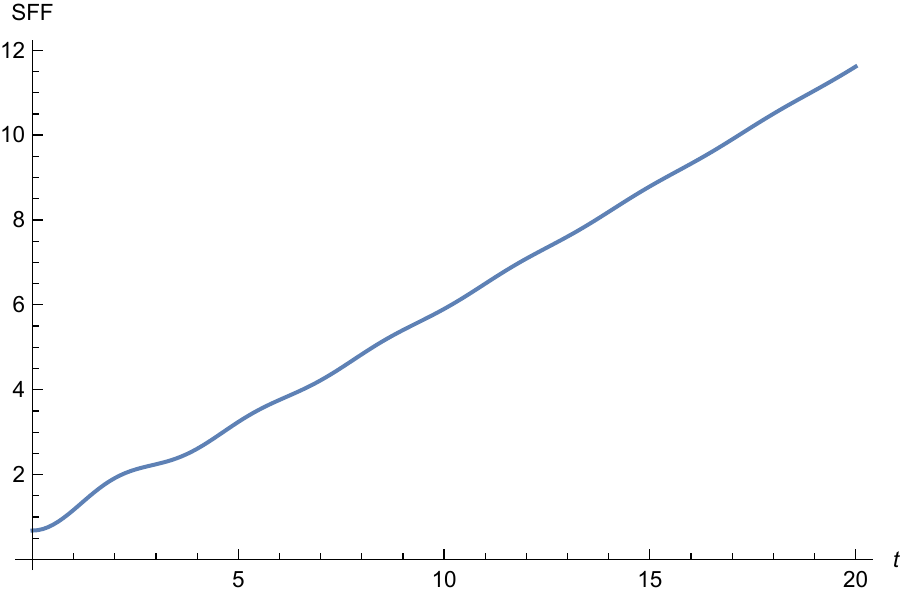}}
\hskip5mm
\subcaptionbox{$\lap=\qu$\label{sfig:S2}}{\includegraphics
[width=0.45\linewidth]{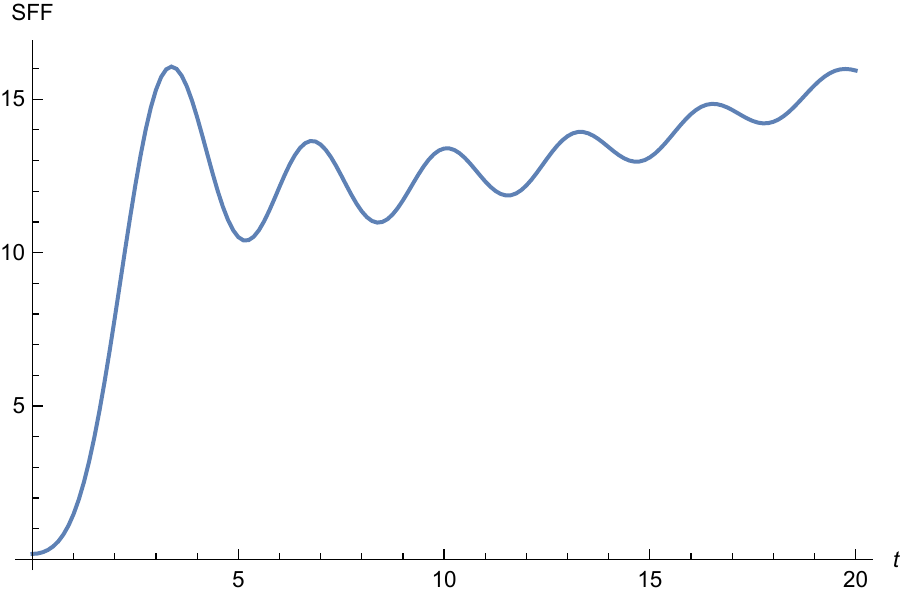}}
  \caption{
Plot 
of the spectral form factor $Z_{\text{cylinder}}(\bt+\ri t,\bt-\ri t,\lap)$
as a function of $t$.
We set $u_0=1,\bt=1,q=1/2$ in \eqref{eq:cyl-matter}
and evaluate it numerically by truncating the summation up to $\fb_{\text{cut}}=100$.
}
  \label{fig:SFF}
\end{figure}

One advantage of DSSYK over the continuum approach of JT gravity is that
one can easily incorporate the effect of matter field within the framework of
chord diagrams. 
According to \cite{Jafferis:2022wez}, if we include the loop correction
of the matter field with dimension $\lap$, the cylinder amplitude
is modified as
\begin{equation}
\begin{aligned}
 Z_{\text{cylinder}}(\bt_1,\bt_2,\lap)&=\sum_{\fb=0}^\infty
\fb I_{\fb}(\bt_1u_0)I_{\fb}(\bt_2u_0)\sum_{n=0}^\infty\left(\frac{q^{\lap\fb}}{1-q^\fb}\right)^n\\
&=\sum_{\fb=0}^\infty
\fb I_{\fb}(\bt_1u_0)I_{\fb}(\bt_2u_0)\frac{1-q^\fb}{1-q^\fb-q^{\lap\fb}}.
\end{aligned} 
\label{eq:cyl-matter}
\end{equation} 
%In the JT gravity case, 
In the semi-classical limit \eqref{eq:semi},
the denominator is replaced by $1-e^{-b}-e^{-\lap b}$
which vanishes at a certain value of $b$. This leads to a divergence 
which was interpreted as a Hagedorn divergence in \cite{Jafferis:2022wez}.
On the other hand, the denominator in \eqref{eq:cyl-matter}
does not vanish for a generic value of $q,\lap$ with $\fb\in\mathbb{Z}_{\geq0}$.
Thus we expect that the Hagedorn divergence from the matter loop is regularized in
DSSYK.
We do not know how to perform the sum over $\fb$ in \eqref{eq:cyl-matter}
in a closed form, but we can easily evaluate \eqref{eq:cyl-matter}
numerically by truncating the summation up to some cut-off $\fb\leq\fb_{\text{cut}}$.
In figure \ref{fig:SFF}, we show the plot of spectral form factor (SFF)
$ Z_{\text{cylinder}}(\bt+\ri t,\bt-\ri t,\lap)$
as a function of $t$.
When $\lap$ is large, the effect of matter loop is negligible and the SFF
exhibits a linear growth called ``ramp'' (see figure \ref{sfig:S1}). When $\lap$ becomes small,
the effect of matter loop drastically changes the behavior of SFF; as we can see from figure 
\ref{sfig:S2}, SFF exhibits an oscillatory behavior as a function of $t$.
It would be interesting to understand the bulk gravitational interpretation of this
oscillation.  

\section{Conclusions and outlook}\label{sec:conclusion}
In this paper, we have studied the EOW brane in DSSYK.
We found that the boundary state of EOW brane is a coherent state
of the $q$-deformed oscillator and the associated
orthogonal polynomial is the continuous big $q$-Hermite polynomial $H_n(x,a|q)$.
We showed that $\psi_n(\th)$ in \eqref{eq:psi-n}  reduces to the 
Whittaker function in the triple scaling limit \eqref{eq:triple}.
We also computed the half-wormhole amplitude in DSSYK
and found the decomposition \eqref{eq:EOWtrumpet-sum} into the trumpet and the factor
$\frac{a^\fb}{1-q^\fb}$,
which is a complete parallel of the decomposition
of half-wormhole amplitude
\eqref{eq:EOWampJT} in JT gravity.

There are many interesting open questions.
Recall that $(L,b)$ and $(n,\fb)$ are related by
(see \eqref{eq:triple} and \eqref{eq:semi})
\begin{equation}
\begin{aligned}
 L=\la n+\log\la,\quad
b=\la\fb.
\end{aligned} 
\end{equation}
Thus the bulk geodesic lengths $L$ and $b$ are quantized in units of $\la$
and there is a minimal length $\la$ in the bulk spacetime.
It would be interesting to understand the physical
meaning of this minimal length. \footnote{In \cite{Berkooz:2022mfk} it is discussed
that the bulk spacetime of DSSYK becomes a non-commutative $AdS_2$
with a minimal length scale $\la$.}

We can generalize our setup by introducing two EOW branes and 
consider the amplitude
\begin{equation}
\begin{aligned}
 \frac{\bra B_a|e^{\bt T}|B_b\ket}{\bra B_a|B_b\ket}=
\int_0^\pi\frac{d\th}{2\pi}\mu(\th)\frac{\bra B_a|\th\ket\bra\th|B_b\ket}{\bra B_a|B_b\ket}e^{-\bt E(\th)}.
\end{aligned} 
\end{equation}
Then the measure factor becomes
\begin{equation}
\begin{aligned}
 \mu(\th)\frac{\bra B_a|\th\ket\bra\th|B_b\ket}{\bra B_a|B_b\ket}=\mu(\th)\frac{(ab;q)_\infty}{(ae^{\pm\ri\th},be^{\pm\ri\th};q)_\infty},
\end{aligned} 
\end{equation}
and the associated orthogonal polynomial is the Al-Salam-Chihara polynomial
$Q_n(x,a,b|q)$. 
This type of boundary condition has been considered
in \cite{sasamoto1999one} in the context of ASEP
(asymmetric simple exclusion process)
under the identification
\begin{equation}
\begin{aligned}
 \bra W|=\bra B_a|,\quad |V\ket=|B_b\ket.
\end{aligned} 
\end{equation}
It would be interesting to pursue this analogy with ASEP further.
One missing ingredient in this analogy is
the relation to the spin chain.
In the context of ASEP, the problem can be mapped to the matrix product state
of open XXZ spin chain and the quantum group naturally arises in the spin 
chain picture (see e.g. \cite{Crampe:2014aoa} for a review).
It would be interesting to understand the role of spin chain in DSSYK, if any.

As discussed in \cite{Jafferis:2022wez}, one can construct various amplitudes
of DSSYK by gluing basic building blocks such as the trumpet, in a similar manner as the JT gravity matrix model \cite{Saad:2019lba}.
However, we still do not have a complete set of the ``Lego block'' of
DSSYK. For instance, we do not know the DSSYK analogue of the Weil-Petersson 
volume, which played an important role for the construction of higher genus amplitudes
of JT gravity.
Also, we have only briefly studied the effect of matter loop in section \ref{sec:trumpet}.
Clearly, more work needs to be done to better understand 
the structure of DSSYK.
%%%%%%%%%%%%%%
\acknowledgments
The author would like to thank Satoru Odake and Kenta Suzuki for discussion.
This work was supported
in part by JSPS Grant-in-Aid for Transformative Research Areas (A) 
``Extreme Universe'' 21H05187 and JSPS KAKENHI Grant 22K03594.

%%%%%%%%%%%%%%%%%%%%%%%%
\appendix

\section{Useful formulae}\label{app:formula}
The $q$-Pochhammar symbol is defined by
\begin{equation}
\begin{aligned}
 (a;q)_n=\prod_{i=0}^{n-1}(1-aq^i).
\end{aligned} 
\end{equation}
We also use the notation
\begin{equation}
\begin{aligned}
 (a_1,\cdots,a_r;q)_n=\prod_{j=1}^r(a_j;q)_n
\end{aligned} 
\end{equation}
and we often use the shorthand notation such as
\begin{equation}
\begin{aligned}
 (ae^{\ri(\pm\th_1\pm\th_2)};q)_\infty=\prod_{s_1,s_2=\pm} (ae^{\ri s_1\th_1+\ri s_2\th_2};q)_\infty.
\end{aligned} 
\end{equation}
The $q$-binomial coefficient is defined by
\begin{equation}
\begin{aligned}
 \qb{n\\k}=\frac{(q;q)_n}{(q;q)_k(q;q)_{n-k}}.
\end{aligned} 
\end{equation}
Below we summarize useful summation formulas
\begin{equation}
\begin{aligned}
 \frac{1}{(t;q)_\infty}&=\sum_{n=0}^\infty\frac{t^n}{(q;q)_n}.
\label{eq:q1}
\end{aligned} 
\end{equation}
\begin{equation}
\begin{aligned}
(t;q)_\infty&=\sum_{n=0}^\infty(-1)^n q^{\hf n(n-1)}\frac{t^n}{(q;q)_n}. 
\end{aligned} 
\end{equation}

\begin{equation}
\begin{aligned}
 (a;q)_n&=\sum_{k=0}^n(-a)^k q^{\hf k(k-1)}\qb{n\\k}.
\end{aligned} 
\end{equation}
\begin{equation}
\begin{aligned}
 \frac{(ta;q)_\infty}{(t;q)_\infty}
&=\sum_{n=0}^\infty(a;q)_n\frac{t^n}{(q;q)_n}.
\end{aligned} 
\end{equation}

\section{Continuous big $q$-Hermite polynomials}\label{app:bigq}
We summarize some useful properties of the 
continuous big $q$-Hermite polynomial $H_n(x,a|q)$.
$H_n(x,a|q)$ is defined recursively by the three-term recurrence relation
\begin{equation}
\begin{aligned}
 2xH_n(x,a|q)=H_{n+1}(x,a|q)+aq^n H_n(x,a|q)+(1-q^n)H_{n-1}(x,a|q),
\end{aligned} 
\end{equation}
with the initial condition $H_{-1}(x,a|q)=0, H_0(x,a|q)=1$.
When $a=0$,  the continuous big $q$-Hermite polynomial
reduces to the $q$-Hermite polynomial
\begin{equation}
\begin{aligned}
 H_n(x,0|q)=H_n(x|q).
\end{aligned} 
\end{equation}
Setting $x=\cos\th$, we can write down the explicit form of 
$H_n(x,a|q)$
\begin{equation}
\begin{aligned}
 H_n(\cos\th,a|q)&=\sum_{k=0}^n\qb{n\\k}
(ae^{\ri\th};q)_ke^{\ri(n-2k)\th}\\
&=\sum_{k=0}^n\qb{n\\k}(-a)^kq^{\hf k(k-1)}H_{n-k}(\cos\th|q).
\end{aligned} 
\label{eq:big-exp}
\end{equation}
When $a=0$, the first line of \eqref{eq:big-exp} reduces to
\begin{equation}
\begin{aligned}
 H_{n}(\cos\th|q)=\sum_{k=0}^n\qb{n\\k}e^{\ri(n-2k)\th}.
\end{aligned} 
\label{eq:Hn-exp}
\end{equation}
An important property of $H_n(x,a|q)$
is the orthogonality relation
\begin{equation}
\begin{aligned}
 \int_0^\pi\frac{d\th}{2\pi}\frac{\mu(\th)}{(ae^{\pm\ri\th};q)_\infty}
H_n(x,a|q)H_m(x,a|q)=(q;q)_n\cob_{n,m},
\end{aligned} 
\end{equation}
where $\mu(\th)=(q,e^{\pm2\ri\th};q)_\infty$.
The generating function
of $H_n(x,a|q)$ is given by
\begin{equation}
\begin{aligned}
 \sum_{n=0}^\infty\frac{t^n}{(q;q)_n}H_n(\cos\th,a|q)=\frac{(ta;q)_\infty}
{(te^{\pm\ri\th};q)_\infty}.
\end{aligned} 
\end{equation}
As a special case of this relation, we find the generation function of
the $q$-Hermite polynomial
\begin{equation}
\begin{aligned}
 \sum_{n=0}^\infty\frac{t^n}{(q;q)_n}H_n(\cos\th|q)=
\frac{1}
{(te^{\pm\ri\th};q)_\infty}.
\end{aligned} 
\label{eq:H-gen}
\end{equation}
Also note that $H_n(x,a|q)$ satisfies a $q$-difference equation for the parameter
$a$ \cite{floreanini1995algebraic}
\begin{equation}
\begin{aligned}
 H_n(x,aq|q)-H_n(x,a|q)=a(1-q^n)H_{n-1}(x,aq|q).
\end{aligned} 
\end{equation}

The product of $q$-Hermite polynomials is expanded as
(see e.g. \cite{ismail1988askey})
\begin{equation}
\begin{aligned}
 H_n(x|q)H_m(x|q)&=\sum_{k=0}^{\min(n,m)}\frac{(q;q)_n(q;q)_m}{(q;q)_{n-k}
(q;q)_{m-k}(q;q)_k}H_{n+m-2k}(x|q),
\end{aligned} 
\label{eq:linearization}
\end{equation}
which is called the linearization formula.

%%%%%%%%%%%%%%%%%%%%%%%%%
\bibliography{paper}
\bibliographystyle{utphys}

\end{document}